\documentclass[12pt,times]{article}
\renewcommand{\baselinestretch}{1.3}

\usepackage{amsmath}
\usepackage{color}
\usepackage{latexsym,amsmath,amssymb}
\usepackage{graphicx,graphics,subfigure}
\usepackage{algorithm,algorithmicx,algpseudocode}

\usepackage[top=0.85in, left=1.0in, bottom=0.9in, right=1.0in]{geometry}

\newtheorem{theorem}{Theorem}

\newtheorem{lemma}{Lemma}
\newtheorem{example}{Example}
\newtheorem{definition}{Definition}

\begin{document}

\title{Space-filling Latin Hypercube Designs based on Randomization Restrictions in Factorial Experiments}

\renewcommand{\baselinestretch}{1.0}
\author{Pritam Ranjan$^*$ and Neil Spencer\\ \small Department of Mathematics and Statistics, Acadia University, Wolfville, Canada, B4P2R6\\ \small $^*$Corresponding author: Pritam Ranjan, email: pritam.ranjan@acadiau.ca \\[-0.5cm]}


\date{}

\maketitle

\begin{abstract}
Latin hypercube designs (LHDs) with space-filling properties are widely used for emulating computer simulators. Over the last three decades, a wide spectrum of LHDs have been proposed with space-filling criteria like minimum correlation among factors, maximin interpoint distance, and orthogonality among the factors via orthogonal arrays (OAs). Projective geometric structures like spreads, covers and stars of $PG(p-1,q)$ can be used to characterize the randomization restriction of multistage factorial experiments. These geometric structures can also be used for constructing OAs and nearly OAs (NOAs). In this paper, we present a new class of space-filling LHDs based on NOAs derived from stars of $PG(p-1, 2)$. \\

\noindent Keywords: $ $ Computer experiments; Nearly orthogonal arrays; Spreads; Stars.

\end{abstract}
\renewcommand{\baselinestretch}{1.3}

\section{Introduction}\label{sec:intro}


Latin hypercube sampling is a statistical method for generating a collections of points from a multi-dimensional distribution, which was first proposed by McKay et al. (1979) as an alternative to random sampling in the Monte Carlo methods for numerically integrating complex multi-dimensional functions. Later on, the Latin hypercube designs (LHDs) became very popular in computer experiments for building statistical metamodels (Santner et al. 2003). Random LHDs can easily be constructed; however, not all are suitable from a modeling viewpoint, for example, if all points are aligned along the main diagonal of the input space (see Section~\ref{sec:randomLHD} for details).

Since replicate runs of a deterministic computer simulator generate identical outputs, it is preferred that the design points (i.e., the set of input locations for running the simulator) are spread out to fill the input space as evenly as possible. Such a design is referred to as a \emph{space-filling} design. In this paper, we discuss space-filling LHDs, a popular class of designs in computer experiments (see Santner et al. (2003); Fang et al. (2006); and Rasmussen and Williams (2006) for an overview). Over the last three decades, a wide spectrum of LHDs have been proposed with different space-filling criteria, for instance, minimum correlation among factors (Iman and Conover 1982), maximin interpoint distance (Morris and Mitchell 1995), and orthogonality among the factors via OAs (Owen 1992; Tang 1993). Definition~1 of Section~2.1 formalizes the definition of an OA. The construction of LHDs with space-filling criteria like maximin distance or minimum correlation often requires computationally intensive search, whereas, OA-based LHDs are easy to construct as long as the OAs exist.

The existence of a desired OA is not always guaranteed, and the construction can also be challenging (Hedayat et al. 1999). OAs can be constructed using a variety of combinatorial objects like linear codes, difference schemes and mutually orthogonal Latin squares. Rains et al. (2002) discussed the existence and construction of OAs using a spread of a finite projective space, $\mathcal{P}=PG(p-1, q)$, and called them \emph{geometric OAs}. The finite projective space $\mathcal{P}=PG(p-1, q)$ is the set of all $p$-dimensional pencils over $GF(q)$, or equivalently, a geometry whose $\{$points, lines, planes, ..., hyperplanes$\}$ are the subspaces of $V_q^p$ of rank $\{1, 2, 3, ... , p-1\}$, where $V_q^p$ is a vector space of rank $p$ over $GF(q)$, and the dimension of a subspace (or flat) of $\mathcal{P}$ is one less than the rank of a subspace of $V_q^p$. This paper focuses on a class of LHDs that are based on projective space over $GF(2)$, i.e., $PG(p-1,2)$.
%
%

Ranjan et al. (2009) established an equivalence between $2^p$ factorial experiments with multiple randomization restrictions and various geometric structures of $PG(p-1,2)$ (e.g., spreads and covers). Here, a point (or pencil) in $PG(p-1,2)$ corresponds to a factorial effect, and a \emph{spread} of $\mathcal{P}$ is a set of disjoint flats of $\mathcal{P}$ that covers all points of $\mathcal{P}$. For example, in a $2^4$ factorial experiment, $\mathcal{P} = \{A, B, AB, C, AC, ..., ABCD\}$ is a $PG(3,2)$, and $\psi = \{S_1 = \{D, BC, BCD\}, S_2=\{C,AB,ABC\}, S_3=\{B,ACD, ABCD\}, S_4 = \{A,BD,ABD\}, S_5=\{CD,AC,AD\} \}$ is a spread of $1$-flats of $\mathcal{P}$. Randomization restrictions at stage $i$ of a factorial experiment is characterized by a \emph{randomization defining contrast subspace} (RDCSS) obtained by spanning $t_i (\le p)$ linearly independent randomization factors (or factorial effects), which is equivalent to a $(t_i-1)$-flat of $\mathcal{P}$ (e.g., $S_i$'s in $\psi$). Such RDCSSs are similar to block defining contrast subgroups in a blocked factorial design, but have to be separate for every stage. See Section~\ref{subsec:review-RDCSS} for a detailed discussion on RDCSSs.

For efficient analysis of a multistage factorial experiment, it is desirable to construct disjoint RDCSSs. However, in many practical situations (e.g., the plutonium alloy experiment of Bingham et al. 2008), overlap among the RDCSSs cannot be avoided. For such cases, Ranjan et al. (2010) proposed designs based on a new geometric structure called a \emph{star} - a set of distinct flats of $PG(p - 1, q)$ that share a common overlap (the nucleus). A star that is also a cover (referred to as a \emph{covering star}) of $PG(p-1,q)$ simplifies to a spread if the nucleus is empty. For example, in a $2^5$ factorial experiment, $\mathcal{P} = \{A, B, AB, C, AC, ..., ABCDE\}$ is a $PG(4,2)$, and $\Omega = \{R_1, R_2, R_3, R_4, R_5\}$ is a covering star with five rays, $R_1 = \langle D, BC, ABCDE\rangle$, $R_2=\langle C,AB,ABCDE\rangle$, $R_3=\langle B,ACD, ABCDE\rangle$, $R_4 = \langle A,BD,ABCDE\rangle$ and $R_5=\langle CD,AC,ABCDE\rangle \}$, and nucleus $\pi = \{ABCDE\}$ of $\mathcal{P}$, where $\langle F_1,...,F_n \rangle$ denotes the span of $F_1,...,F_n$.


We have discovered a new class of space-filling LHDs that can be constructed using stars of $PG(p-1, 2)$. It turns out that a star with non empty nucleus generates near-OAs (NOAs). In general, a near-OA is an array in which the orthogonality requirement is nearly satisfied (for details, see Taguchi 1959; Wang and Wu 1992; Nguyen 1996; Wu and Hamada 2000; and Xu 2002). In the spirit of Rains et al. (2002), we sometimes refer to these star-based NOAs as \emph{geometric NOAs}. By following Tang's OA-based LHD construction algorithm, we construct a class of geometric NOA-based LHDs. Although such LHDs are not always very space-filling, a near orthogonality (e.g., Xu and Wu 2001) or space-filling criterion can be used to search for a good one. To avoid the search, we also propose a set of guidelines for carefully distributing the factorial effects among RDCSSs of the star which ensures space-filling LHDs. It is worth noting that the existence of OA-based LHDs are limited to only few $n \times d$ combinations, whereas, the existence conditions for stars are less stringent.

The remainder of the paper is organized as follows. Section~2 presents an overview of LHDs, RDCSSs in a $2^p$ factorial experiment, and spreads and stars of a $PG(p-1, q)$. In Section~3, we establish theoretical results for the existence and an algorithm of the construction of geometric-NOAs. Section~4 concludes the paper with a few remarks.

\section{Background}\label{sec:background}


This section starts with a brief review on random LHDs and OA-based LHDs. Then a few results are presented to establish the equivalence between a multistage factorial design with randomization restrictions and geometric structures of $PG(p-1, 2)$.

\subsection{Latin hypercube designs}\label{sec:randomLHD}

Let $L(n, d)$ be an LHD with $n$ runs in $d$ factors (dimension of input space), where $L_{ij}$ denotes the level of factor $j$ in the $i$-th experimental run, and each factor includes $n$ uniformly spaced levels.
%
%
%
In computer experiments, the input spaces are typically bounded hyper-rectangles, and can be transformed to unit hypercubes. A random $L(n, d)$ in $[0, 1]^d$ has $L_{ij} = (\pi_j(i) - u_{ij})/n$ for $1 \le j \le d$ and $1 \le i \le n$, where $u_{ij} \sim Unif(0,1)$ and $(\pi_j(1), ..., \pi_j(n))$ is a random permutation of $\{1, ..., n\}$ (see Tang (1993) for details). Ignoring the $Unif(0,1)$ perturbations, there are $(n!)^d$ distinct LHDs.

\begin{figure}[h!]\centering
\subfigure[A good design for modeling]{\label{fig:good_random_LHD} \includegraphics[width=3in]{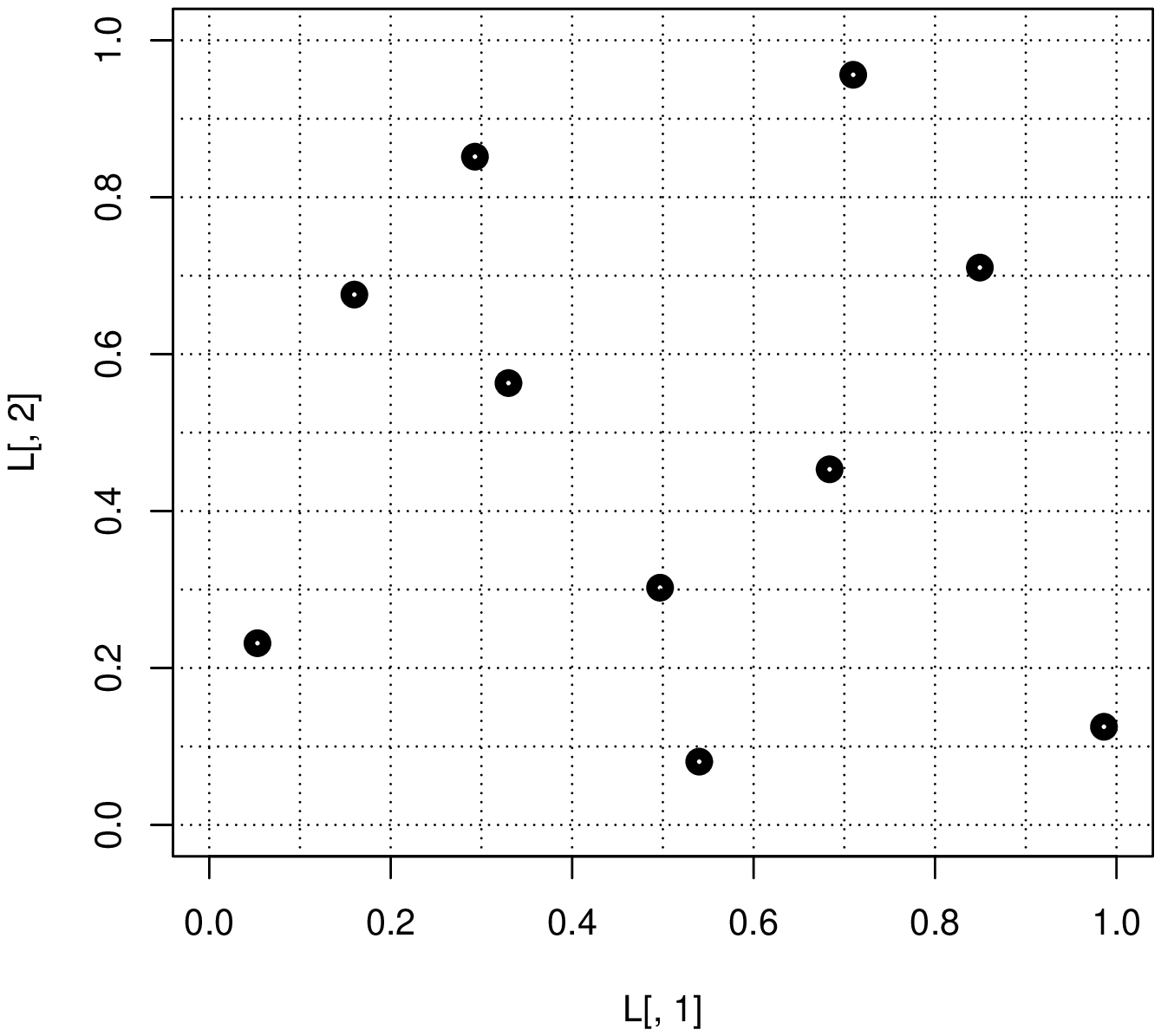}}
\subfigure[A bad design for modeling]{\label{fig:bad_random_LHD} \includegraphics[width=3in]{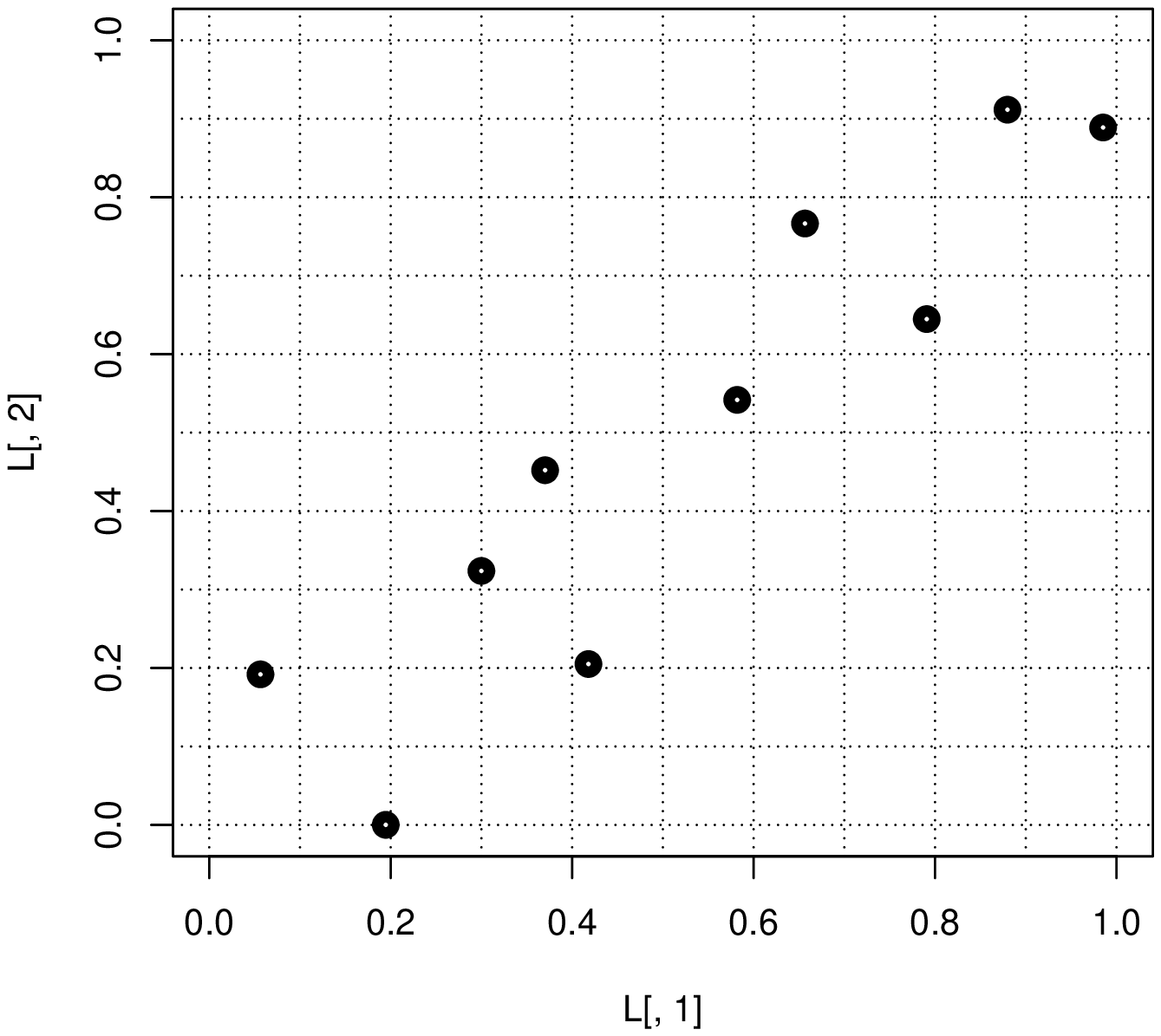}}\vspace{-0.2cm}
\caption{Two realizations of random LHDs in $[0, 1]^2$.}\label{fig:random_LHDs}
\end{figure}

Although LHDs have a nice one-dimensional projection property, that is, one point each in $((i-1)/n, i/n)$ for $1 \le i \le n$, random LHDs can be quite undesirable from a modeling viewpoint. Figure~\ref{fig:random_LHDs} presents two realizations of random LHDs in $[0, 1]^2$. The points in Figure~\ref{fig:good_random_LHD} are distributed throughout the whole space (space-filling), but the points in Figure~\ref{fig:bad_random_LHD} are concentrated along the main diagonal.

In this paper, we propose a class of LHDs, called star-based LHDs, which are space-filling under certain conditions. Star-based LHDs are generalizations of the OA-based LHDs.

\begin{definition}\label{def:OA}
An $n\times d$ array $\mathcal{A}$ denoted by $OA(n, s_1s_2 \cdots s_d, r)$ is said to be a strength $r$ OA with $n$ runs and $d$ factors, if factor $j$ has $s_j$ levels $\{0, ... ,s_j-1\}$ and each $n \times r$ subarray contains every possible $r$-tuple an equal number of times.
\end{definition}

The special case of $s_1=s_2=\cdots=s_d$ corresponds to a symmetric OA denoted by $OA(n, s, d, r)$. Although mixed-level (or asymmetric) OAs have been investigated in recent years (Hedayat et al. 1999)), it is a less explored area than symmetric OAs.

A simple existence condition of an asymmetric OA of strength $r$ follows from the strength aspect of Definition~\ref{def:OA}, that is, $n$ must be a multiple of $s_1^{x_1}s_2^{x_2}\cdots s_d^{x_d}$ for every set of $x_1,...,x_d \in \{0,1\}$ such that $\sum_{j=1}^d x_j \le r$. Another popular existence result comes from the Rao bound: $n-1 \ge \sum_{j=1}^d (s_j-1)$. Despite these results, determining the existence of a desired OA is nontrivial, and the difficulty increase as the strength $r$ and the number of levels $s_j$ increase. Even OAs with strength $2$ do not always exist for arbitrary $n$ and $d$ (see Hedayat et al. (1999) and Rains et al. (2002) for more results).

An $OA(n, s_1s_2 \cdots s_d, r)$-based LHD is constructed in two steps (Tang 1993). First the OA (say $\mathcal{A}$) is used to construct an array (say $\mathcal{L}$) by replacing the $n/s_j$ entries of the $j$th column with value $k$ by a random permutation of $(k-1)n/s_j + 1, (k-1)n/s_j + 2, ..., kn/s_j$, for $k=1,...,s_j$ and $j=1,...,d$. Then, the desired OA-based LHD, $L(n,d)$, is given by $L_{ij}=(\mathcal{L}_{ij} - u_{ij})/n$ for $i=1,...,n$, $j=1,...,d$ and $u_{ij} \sim Unif(0,1)$. Note that the randomness in an OA-based LHD is introduced via the random permutation of $(k-1)n/s_j + 1, (k-1)n/s_j + 2, ..., kn/s_j$ and the uniform perturbation $u_{ij}$.

\begin{figure}[h!]\centering
\includegraphics[width=6in]{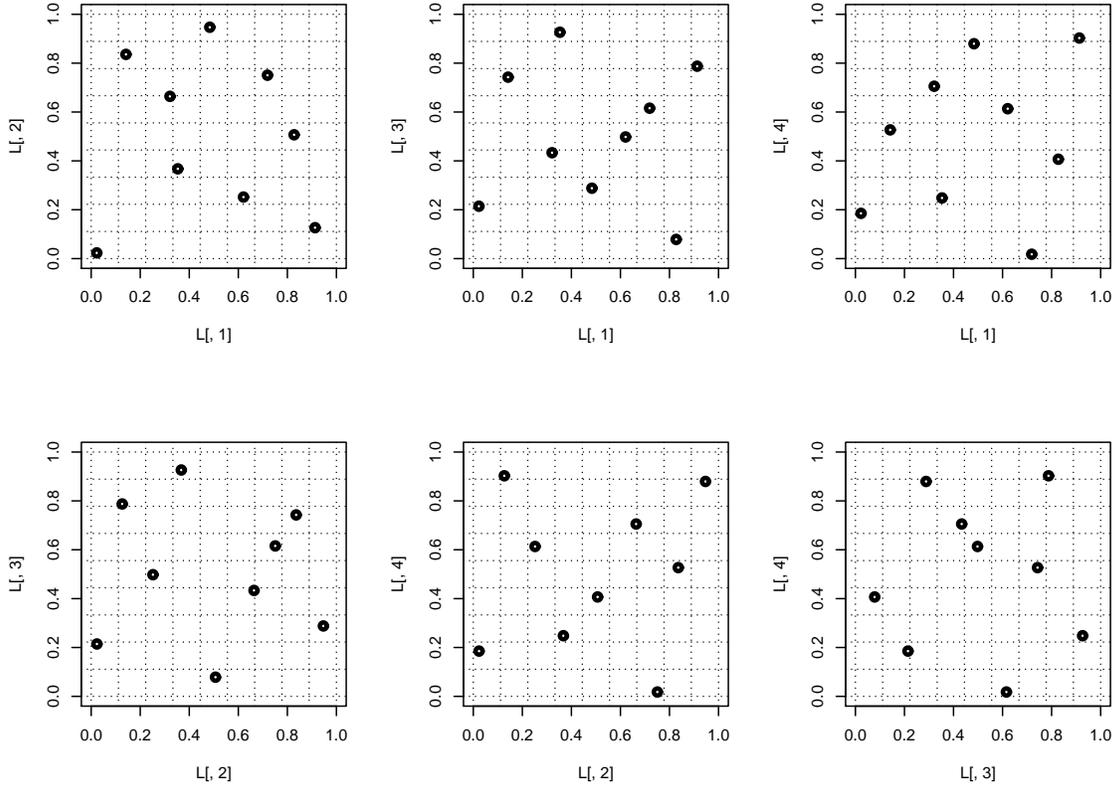}\vspace{-0.5cm}
\caption{Two dimensional projections of a random $OA(9,3,4,2)$ -based LHD in $[0,1]^4$.}\label{fig:OA9432-LHD-L1}
\end{figure}

For example, an $OA(9,3,4,2)$ given by
$$
 \mathbf{\mathcal{A}^T} =\left( \begin{array}{ccccccccc}
  0  &  0 &   0  &  1 &   1  &  1  &  2  &  2  & 2\\[-0.1cm]
  0  &  1 &   2  &  0 &   1  &  2  &  0  &  1  & 2\\[-0.1cm]
  0  &  1 &   2  &  1 &   2  &  0  &  2  &  0  & 1\\[-0.1cm]
  0  &  2 &   1  &  1 &   0  &  2  &  2  &  1  & 0\\
 \end{array}
 \right),$$
with a random permutation of $(k-1)n/s_j + 1, (k-1)n/s_j + 2, ..., kn/s_j$ generates
$$
 \mathbf{\mathcal{L}^{T}} =\left( \begin{array}{ccccccccc}
 1  &  3 &   2  &  6  &  4  &  5 &   9 &   8  &  7\\[-0.1cm]
 1  &  6 &   8  &  3  &  4  &  9 &   2 &   5  &  7\\[-0.1cm]
 2  &  4 &   7  &  5  &  9  &  3 &   8 &   1  &  6\\[-0.1cm]
 2  &  7 &   5  &  6  &  3  &  8 &   9 &   4  &  1\\
 \end{array}
 \right),
 $$
and the corresponding LHD is shown in Figure~\ref{fig:OA9432-LHD-L1}.
%

%
%

In this paper, we propose a new class of star-based geometric NOAs for constructing space-filling LHDs. The results in the next section show that the existence conditions for such NOAs are less stringent as compared to OAs.

\subsection{RDCSSs and Projective Geometries}\label{subsec:review-RDCSS}

Ranjan et al. (2009) proposed a unified framework using finite projective geometry for the existence and construction of factorial designs with randomization restriction (e.g., nested designs, split-plot designs, split-lot designs, and combinations thereof). For a $2^p$ factorial experiment, $\mathcal{P}=PG(p-1,2)$ denotes the set of all $2^p-1$ factorial effects (excluding the grand mean), and a $p$-dimensional pencil of $\mathcal{P}$ (or equivalently, a vector in $V_2^p$) with $r\, (\le p)$ nonzero elements uniquely corresponds to an $r$-factor interaction.


The restrictions on the randomization of experimental runs lead to grouping experimental units into sets of trials. These sets are formed using linearly independent pencils (points or effects) of $\mathcal{P}$, also referred to as the \emph{randomization restriction factors} (like the blocking factors in a blocked factorial design). A set $S$ of all non-null pencils formed by linear combinations of $t$ independent randomization restriction factors in $\mathcal{P}$ constitutes a $(t - 1)$-dimensional subspace (or, $(t-1)$-flat) of $\mathcal{P}$ with $|S|=2^t-1$. We call such a subspace a $t$-dimensional randomization defining contrast subspace (RDCSS). For example, $S_1 = \{A, BD, ABD\}$ and $S_2=\{B, ACD, ABCD\}$ are $2$-dimensional RDCSSs in a $2^4$ experiment, where $A, BD$ and $B, ACD$ are linearly independent randomization restriction factors for $S_1$ and $S_2$.

For efficient analysis of multistage factorial experiments, it is desirable to construct disjoint RDCSSs. Ranjan et al. (2009) established the existence of a set of disjoint RDCSSs in a $2^p$ factorial experiment via the existence of a \emph{spread} of a $PG(p-1, 2)$.

\begin{definition}\label{def1}
For $1 \le t \le p$, a balanced $(t-1)$-spread of $\mathcal{P}=PG(p-1,q)$ is a set, $\psi$, of $(t-1)$-flats of $\mathcal{P}$ which partitions $\mathcal{P}$.
\end{definition}

The size of a balanced $(t-1)$-spread $\psi$ of $PG(p-1,q)$ is $|\psi| = (q^p-1)/(q^t-1)$. A necessary and sufficient condition for the existence of a $(t-1)$-spread is that $t$ divides $p$ (Andr\'e 1954). See Ranjan et al. (2009) for more results on the balanced and mixed (partial) spreads. For cases in which overlap cannot be avoided, Ranjan et al. (2010) proposed designs based on a new geometric structure called a \emph{star} - a set of distinct flats of $PG(p - 1, q)$ that share a common overlap (the nucleus). A star of $\mathcal{P}$ which spans all points of $\mathcal{P}$ is referred to as a covering star of $\mathcal{P}$. For example, $\Omega = \{R_1, R_2, R_3, R_4, R_5\}$ is a balanced covering star of $PG(4,2)$ with five $2$-flat rays and $0$-flat nucleus $\pi = \{ABCDE\}$.

\begin{definition}\label{def:star}
A balanced covering star $\Omega = St(\mu, t, t_0)$ of $\mathcal{P} = PG(p-1,q)$ is a set of $\mu$ rays ($(t-1)$- flats) and a nucleus ($(t_0-1)$- flat), where the nucleus is contained in each of the $\mu$ rays (i.e., $t_0 < t$), and $\mu = (q^{p-t_0}-1)/(q^{t-t_0}-1)$ .
\end{definition}

A necessary and sufficient condition for the existence of a balanced covering star $St(\mu, t, t_0)$ of $PG(p-1,q)$ is $(t-t_0)$ divides $(p-t_0)$. Let $St(t_1,...,t_{\mu}; t_0)$ be a mixed/unbalanced covering star of $PG(p-1,q)$ with $\mu$ rays and a $(t_0-1)$-dimensional nucleus, such that $t_1\le \cdots \le t_{\mu}$. The next two lemmas are taken from Ranjan et al. (2010).

\begin{lemma}\label{lemma:ns_unbalanced_star}
For the existence of a covering star $St(t_1,...,t_{\mu}; t_0)$ of $\mathcal{P} = PG(p-1,q)$, the following conditions are necessary:
\begin{enumerate}
\item[(i)] $q^{p-{t_0}}-1 = \sum_{i=1}^{\mu} (q^{t_i-{t_0}}-1)$,\\[-1cm]
\item[(ii)] $t_i+t_j-{t_0} \le p$ for every $i \ne j$ $(i,j=1,..., k)$,
\end{enumerate}
\end{lemma}

A necessary and sufficient condition for unbalanced covering star is still unknown. Lemma~\ref{lemma:star_balanced} is a powerful result and guarantees the existence of a balanced star for every $t$ and $p$ ($t<p$).

\begin{lemma}\label{lemma:star_balanced}
For every $t$ $(2\le t <p)$ and ${t_0}=t-1$, there exists a balanced covering
star $St(\mu,t, t_0)$ of $\mathcal{P}=PG(p-1,q)$ with $\mu = (q^{p-t+1}-1)/(q-1)$.
\end{lemma}

Next we discuss how these star-based multistage factorial designs can be used to construct space-filling LHDs. Though we focus on LHDs generated using ``two-level" factorial designs, most of the results can be generalized for $q$-level designs.

\section{Star based LHD}\label{sec:new_material}

First we generalize the spread to OA construction algorithm of Rains et al. (2002) for stars (Algorithm~\ref{algo:NOA_const}). It turns out that the arrays obtained via Algorithm~\ref{algo:NOA_const} are nearly orthogonal if the generating star is non-trivial (i.e., the nucleus is non-empty). We then follow Tang's OA-based LHD construction algorithm (outlined in Section~\ref{sec:randomLHD}) on these arrays for constructing star induced NOA-based LHDs. We also establish new existence results for such geometric NOAs that are derived from stars. Finally, we present a few guidelines for constructing specific star induced NOAs that lead to space-filling LHDs.

We use the binary (vector or pencil) representation of all effects for our construction method. Let $(a_1, a_2, ..., a_{2^p-1})$ be the ordered set of all effects in $\mathcal{P}$, and $St(t_1,...,t_{\mu}; t_0)$ be a covering star of $\mathcal{P}$ with $\mu$ rays $\{R_1, ...,R_{\mu}\}$, where $|R_j|=2^{t_j}-1$ (let $s_j = 2^{t_j}$). Algorithm~\ref{algo:NOA_const} constructs an $NOA(2^p, s_1s_2\cdots s_{\mu}, 2)$ denoted by $\mathcal{A}=[\mathcal{A}_{*1}:\mathcal{A}_{*2}:\cdots:\mathcal{A}_{*\mu}]$.

\begin{algorithm*}
\caption{Star to NOA construction} \label{algo:NOA_const}
\begin{algorithmic}[1]
\For{$j = 1 \to \mu$}
   \State For $j$-th stage of randomization restriction, find $t_j$ linearly independent randomization restriction factors $\{\delta^{(j)}_1 , ..., \delta^{(j)}_{t_j}\}$ such that $R_j = \langle \delta^{(j)}_1 , ..., \delta^{(j)}_{t_j}\rangle $.
      \For{$i = 1 \to 2^p-1$}
          \For{$l = 1 \to t_j$}
             \State Compute $b_{lj}^{(i)} = a_i \cdot \delta_{l}^{(j)}$ \ \ \# inner product over $mod(2)$
          \EndFor
          \State Define $\mathcal{A}_{ij} = \sum_{l=1}^{t_j} b_{lj}^{(i)} 2^{t_j-l}$.
      \EndFor
\EndFor
\end{algorithmic}
\end{algorithm*}


For every $j \in \{1,2,...,\mu\}$, the $i$-th element of the NOA is $\mathcal{A}_{ij} \in \{0, 1, ..., 2^{t_j}-1\}$ for all $1\le i \le 2^p-1$. As a convention, we append a row of zeros at the beginning. See Example~\ref{example:1} for an illustration. This construction ensures the existences of NOAs conditional on the existence of the stars. Next we formalize the existence results for star-based NOAs.

\begin{theorem}\label{them:NOA_general_exist}
The existence of a covering star $St(t_1,...,t_{\mu}; t_0)$ of $PG(p-1, 2)$, is a sufficient condition for the existence of an $NOA(2^p, s_{i_1}s_{i_2}\cdots s_{i_k}, 2)$ with $1\le k \le \mu$, $1\le {i_1} < i_2 < \cdots < i_k \le \mu$ and  $s_j = 2^{t_j}$ for all $1\le j \le \mu$.
\end{theorem}

Similar to Theorem~\ref{them:NOA_general_exist}, the existence of a balanced star $St(\mu; t; t_0)$ of $PG(p-1,2)$ suffices the existence of $NOA(2^p, 2^t, k, 2)$ for all $1\le k\le \mu$. Moreover, it turns out that the maximal $NOA(2^p, 2^t, \mu, 2)$ is a $2^{-t_0}$ fraction of an $OA(2^{p+t_0}, 2^t, \mu, 2)$, and its existence requires the divisibility condition $(t-t_0)|(p-t_0)$ which follows from the existence of a balanced star.

\begin{theorem}\label{them:NOA_balanced_exist}
For every $1\le t < p$, there exist $NOA(2^p, 2^t, k, 2)$ for all $1 \le k \le 2^{p-t+1}-1$, and the maximal $NOA(2^p, 2^t, 2^{p-t+1}-1, 2)$ is a $2^{-(t-1)}$ fraction of an $OA(2^{p+t-1}, 2^t, 2^{p-t+1}-1, 2)$.
\end{theorem}

Proofs of Theorems~\ref{them:NOA_general_exist} and \ref{them:NOA_balanced_exist} follow from Lemmas~\ref{lemma:ns_unbalanced_star} and \ref{lemma:star_balanced}, and the NOA construction in Algorithm~\ref{algo:NOA_const}. The following example illustrates the theoretical results and Algorithm~\ref{algo:NOA_const}.

\begin{example}\label{example:1}{\rm
Suppose $\Omega$ is a covering star of $PG(3,2)$ (or, in a $2^4$ experiment) such that the rays (or RDCSSs) are of size seven each (i.e., $p=4$ and $t=3$). Here, $\mathcal{P}=(a_1,...,a_{15})=(D,C,CD,B,...,ABCD)$ (in binary representation). Since there always exists a balanced covering star for $t_0=t-1$ with $\mu = 2^{p-t+1}-1$ (here $\mu =3$), Theorem~\ref{them:NOA_balanced_exist} ensures the existence of $NOA(16, 8, k, 2)$ for all $1\le k \le 3$. Following the construction in Ranjan et al. (2010), one possible option for $\Omega = \{R_1, R_2, R_3\}$ is $R_1 =\langle A,B,ACD \rangle$, $R_2 =\langle C,D,ABC \rangle$ and $R_3 =\langle AC,BC,AD \rangle$. Note that the nucleus is $\langle AB,CD\rangle$.

The first column of the maximal NOA is $\mathcal{A}_{*1}=(0,\mathcal{A}_{11},\mathcal{A}_{21},...,\mathcal{A}_{15,1})^T$, with, $\mathcal{A}_{i1} = b_{11}^{(i)}2^{3-1} + b_{21}^{(i)}2^{3-2} + b_{31}^{(i)}2^{3-3}$. As an example, for $i=2$, $a_2 = C$ and $b_{11}^{(2)} = (1,0,0,0)^T\cdot (0,0,1,0) = 0$, $b_{21}^{(2)} = (0,1,0,0)^T\cdot (0,0,1,0) = 0$ and $b_{31}^{(2)} = (1,0,1,1)^T\cdot (0,0,1,0) = 1$. Therefore, $\mathcal{A}_{21} = 0\cdot 4 + 0\cdot2 + 1\cdot1=1$. The final $NOA(16,8,3,2)$ is
{\small
$$ \mathbf{\mathcal{A}^T} =\left( \begin{array}{cccccccccccccccc}
0 & 1 & 1 & 0 & 2 & 3 & 3 & 2 & 5 & 4 & 4 & 5 & 7 & 6 & 6 & 7 \\
0 & 2 & 5 & 7 & 1 & 3 & 4 & 6 & 1 & 3 & 4 & 6 & 0 & 2 & 5 & 7 \\
0 & 1 & 6 & 7 & 2 & 3 & 4 & 5 & 5 & 4 & 3 & 2 & 7 & 6 & 1 & 0 \\
      \end{array}
             \right)$$
}
with a corresponding random Latin hypercube array (randomness is introduced via random permutation of labeling) given by
{\small
$$ \mathbf{\mathcal{L}^T} =\left( \begin{array}{cccccccccccccccc}
 1 &   4  &  3 &   2 &   5  &  8 &   7 &   6 &  11 &   10 &    9 &   12 &   16  &  14 &   13 &   15\\
 2 &   6  & 11 &  15 &   3  &  8 &   9 &  14 &   4 &    7 &   10 &   13 &    1  &   5 &   12 &   16\\
 1 &   4  & 14 &  15 &   6  &  8 &  10 &  12 &  11 &    9 &    7 &    5 &   16  &  13 &    3 &    2\\
      \end{array}
             \right),$$
}
and a resultant random LHD (with random uniform perturbation) is shown in Figure~\ref{fig:NOA(16,8,3,2)LHD}.
\begin{figure}[h!]\centering
\includegraphics[height=2.25in,width=6in]{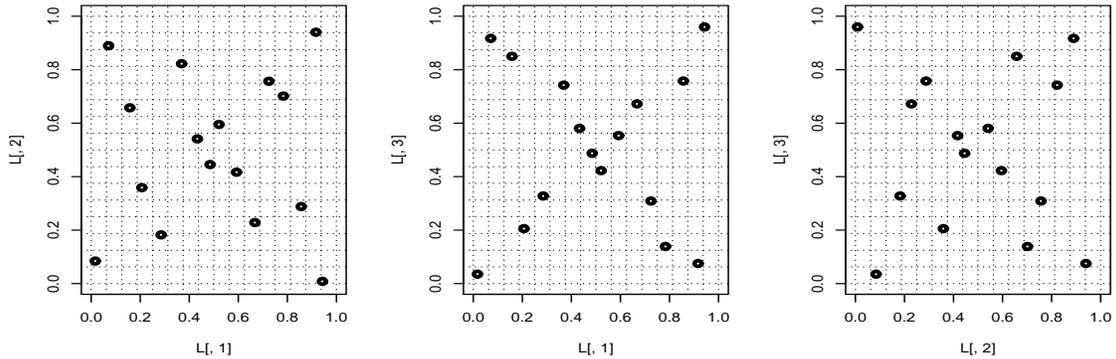}\vspace{-0.5cm}
\caption{Two-dimensional projections of a random $NOA(16,8,3,2)$ -based LHD in $[0,1]^3$.}\label{fig:NOA(16,8,3,2)LHD}
\end{figure}
It is clear from Figure~\ref{fig:NOA(16,8,3,2)LHD} that the two-dimensional projections in the left and right panels appear to be somewhat space-filling; however, the middle panel exhibit specific systematic pattern along the two diagonals and the design points are clearly not space-filling.


}\end{example}

In general, LHDs derived from star-based NOAs have nice geometric features; however, not all such LHDs are very space-filling. One possibility is to use a near orthogonality (Xu and Wu 2001) or space-filling criterion to sort through a set of randomly generated star induced NOA-based LHDs, which may still be computationally intensive. It turns out that the choice of generators (linearly independent randomization factors, $\delta_l^{(j)}$) of rays plays an important role in the geometry of the resultant LHD. Consequently, we propose a few guidelines for carefully choosing and distributing $\delta_l^{(j)}$ in the star to NOA construction (in Algorithm~\ref{algo:NOA_const}) which empowers the space-filling property of LHDs.

The suggested guidelines are as follows:
\begin{itemize}
\item (G1) $\delta_1^{(j)}$ (the first randomization restriction factor) should not be an element of the nucleus for all $1\le j \le \mu$, \\[-0.5cm]
\item (G2) $\delta_l^{(j_1)} \ne \delta_l^{(j_2)}$ ($l$-th generators of $R_{j_1}$ and $R_{j_2}$ should be different) for all $l$ and $1 \le j_1 < j_2 \le \mu$, \\[-0.5cm]
\item (G3) $\delta_{l_1}^{(j_1)} + \delta_{l_2}^{(j_1)} \ne \delta_{l_1}^{(j_2)} + \delta_{l_2}^{(j_2)}$ (the interaction of $l_1$-th and $l_2$-th generators of $R_{j_1}$ should be different than that of $R_{j_2}$) for all $l_1 \ne l_2$ and $1 \le j_1 < j_2 \le \mu$.
\end{itemize}

In the next few examples we use the same star as in Example~1 but carefully choose $\delta_l^{(j)}$'s (of Step~2 in Algorithm~1) for illustrating the importance of these guidelines. The degree to which a design is space-filling can be measured using criteria such as total pairwise correlation or distance based measures. For every LHD $L(n,d)$, we compute both \emph{minimum interpoint distance} (MID) and \emph{average interpoint distance} (AID) among the design points:
$$ MID(L) = \min\{\|L_{i*}-L_{j*}\|, \, 1\le i<j \le n \},$$
$$ AID(L) = \frac{1}{n(n-1)/2}\sum_{i=1}^{n-1}\sum_{j=i+1}^n \|L_{i*}-L_{j*}\|,$$
where $L_{i*}$ is the $i$-th row of the $n\times d$ LHD array $L$, and $\|\cdot\|$ denotes the Euclidean norm. MID guards against the worst case scenario (i.e., the criterion penalizes even if there is only one pair of design points that are close together), whereas AID measures the overall closeness of the design points. For obtaining space-filling LHD we would like to \emph{maximize} both MID and AID values. Of course, the design ranking can be different under different criterion. Although we are primarily interested in ranking the full NOA-based LHDs induced from stars, we also compare the two dimensional projections to highlight the geometric anomalies that occur due to violating the guidelines.

\begin{example}\label{example:2}{\rm
Consider the same star as in Example~\ref{example:1}, that is, a covering star $\Omega = St(3, 3, 2)$ of $PG(3,2)$ (with $p=4$, $t=3$ and $t_0=2$), but choose the generators $\delta_l^{(j)}$'s such that only ``G1" is violated. For instance, let $R_1 =\langle AB,B,ACD \rangle$, $R_2 =\langle D,C,ABC \rangle$ and $R_3 =\langle AC,BC,CD \rangle$. Then the nucleus is $\langle AB, CD\rangle$, and $\delta_1^{(1)} = AB$ belongs to the nucleus. MID and AID values for the full three-dimensional LHD are 0.1875 and 0.6896 respectively. To highlight the geometric structure of the points, Figure~\ref{fig:NOA_LHDs_G1} depicts the two-dimensional projections of the LHD without $Unif(0,1)$ perturbation. 

\begin{figure}[h!]\centering
\subfigure[A bad design ($R_1$ and $R_2$)]{\label{fig:NOA_LHD_G1a} \includegraphics[width=2.0in]{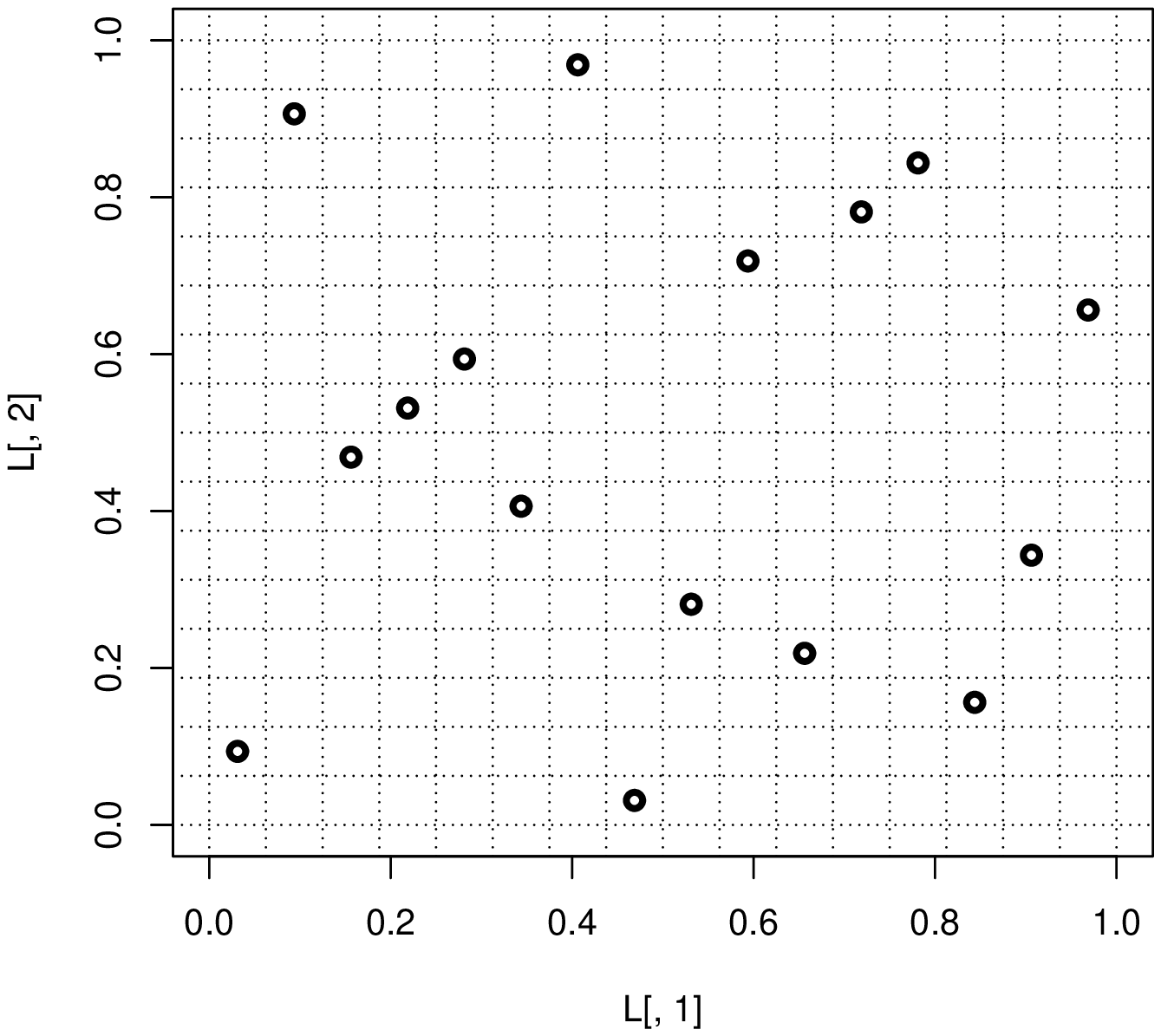}}
\subfigure[A bad design ($R_1$ and $R_3$)]{\label{fig:NOA_LHD_G1b} \includegraphics[width=2.0in]{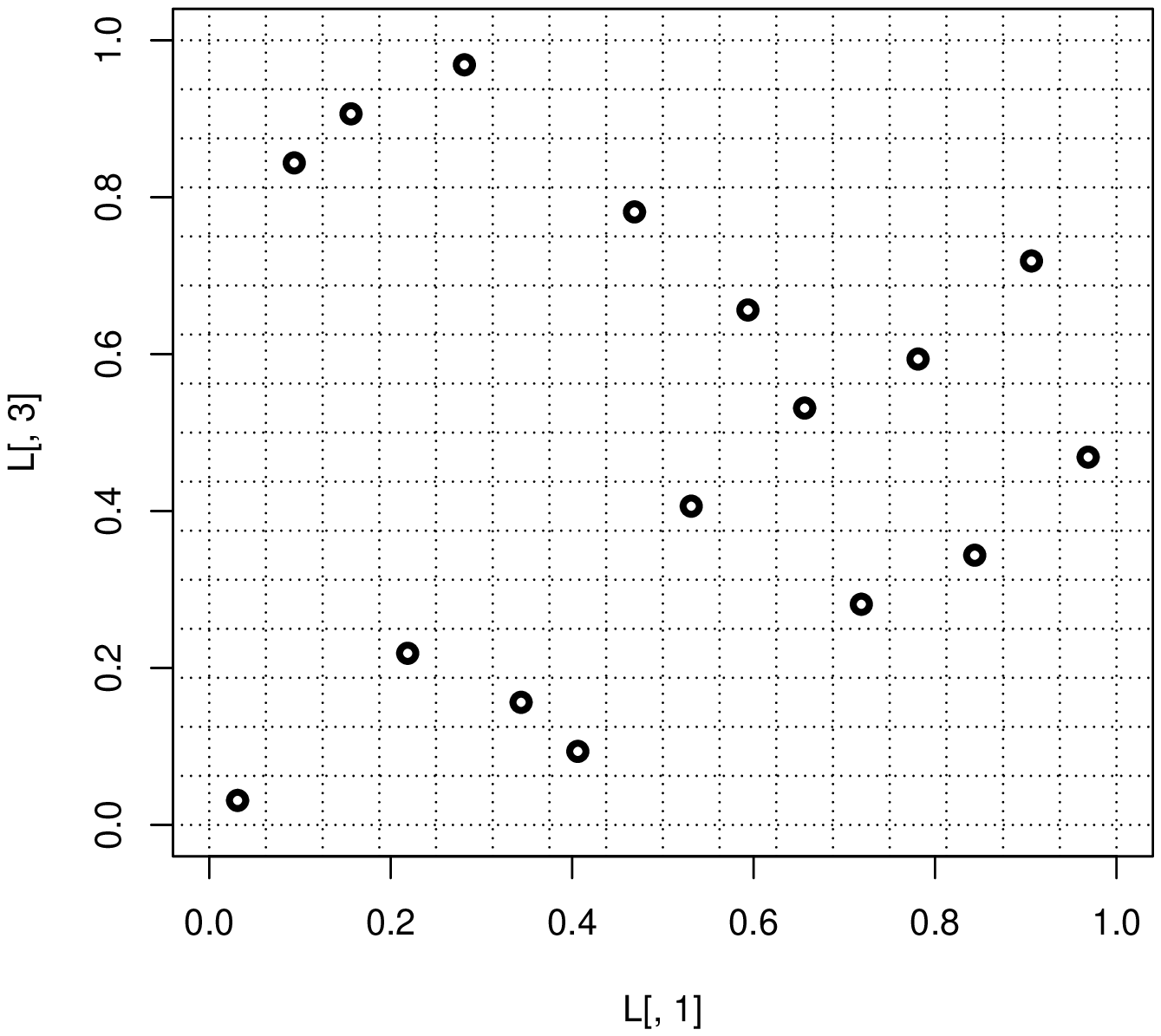}}
\subfigure[A good design ($R_2$ and $R_3$)]{\label{fig:NOA_LHD_G1c} \includegraphics[width=2.0in]{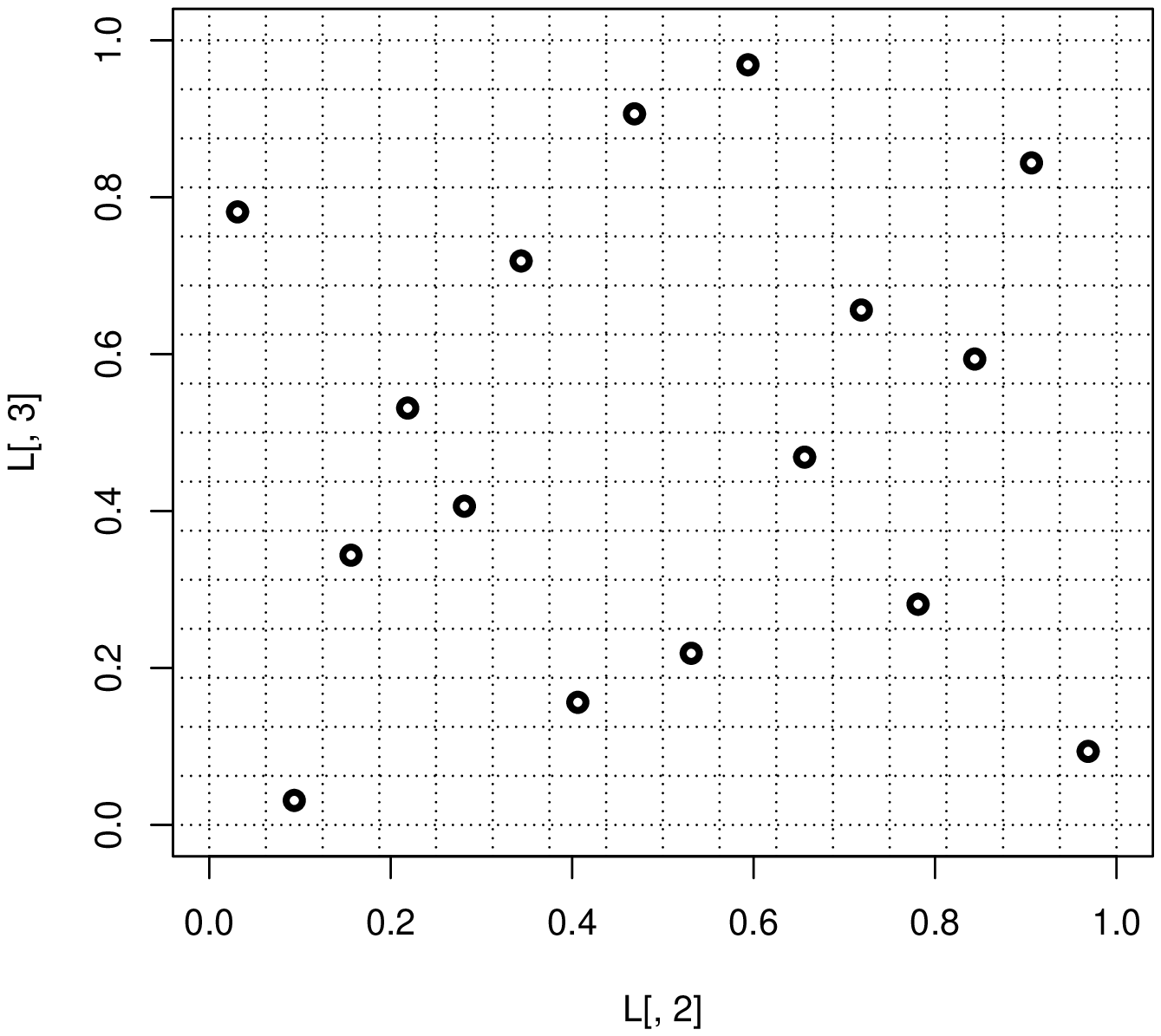}}\vspace{-0.1cm}
\caption{Two-dimensional projections of $NOA(16,8,3,2)$-based LHD in $[0, 1]^3$.}\label{fig:NOA_LHDs_G1}
\end{figure}

Figures~\ref{fig:NOA_LHD_G1a} and \ref{fig:NOA_LHD_G1b} exhibit specific patterns with big holes (particularly Figure~\ref{fig:NOA_LHD_G1b}), whereas Figure~\ref{fig:NOA_LHD_G1c} shows scatter of design points more evenly throughout the design space. MID values for the two-dimensional LHDs in Figures~\ref{fig:NOA_LHD_G1a} and \ref{fig:NOA_LHD_G1b} are same (0.08839) and smaller than the LHD in Figure~\ref{fig:NOA_LHD_G1c} (0.13975). Although AID values do not show strong correlation with MID values, Figure~\ref{fig:NOA_LHD_G1b} yields the smallest AID value (0.5405) and Figures~\ref{fig:NOA_LHD_G1a} and \ref{fig:NOA_LHD_G1c} show comparable AID values (0.5483 and 0.5480). That is, one may argue that the LHD projections with $R_1$ (which violated ``G1") are less space-filling than the projection without $R_1$.
}\end{example}

\begin{example}\label{example:3}{\rm
Consider the same star as in Examples~\ref{example:1} and \ref{example:2}, however, choose $\delta_l^{(j)}$'s such that only ``G2" is violated. Let $R_1 =\langle A,B,ABCD \rangle$, $R_2 =\langle C,D,ABCD \rangle$ and $R_3 =\langle AC,BD,BC \rangle$. Note $\delta_3^{(1)} = \delta_3^{(2)} = ABCD$, which also belongs to the nucleus. MID and AID values for the full three-dimensional LHD are 0.1875 and 0.6867 respectively, that are similar to the design in Example~\ref{example:2}. Figure~\ref{fig:NOA_LHDs_G2} shows the two-dimensional projections. 

\begin{figure}[h!]\centering
\subfigure[An okay design ($R_1$ and $R_2$)]{\label{fig:NOA_LHD_G2a} \includegraphics[width=2.0in]{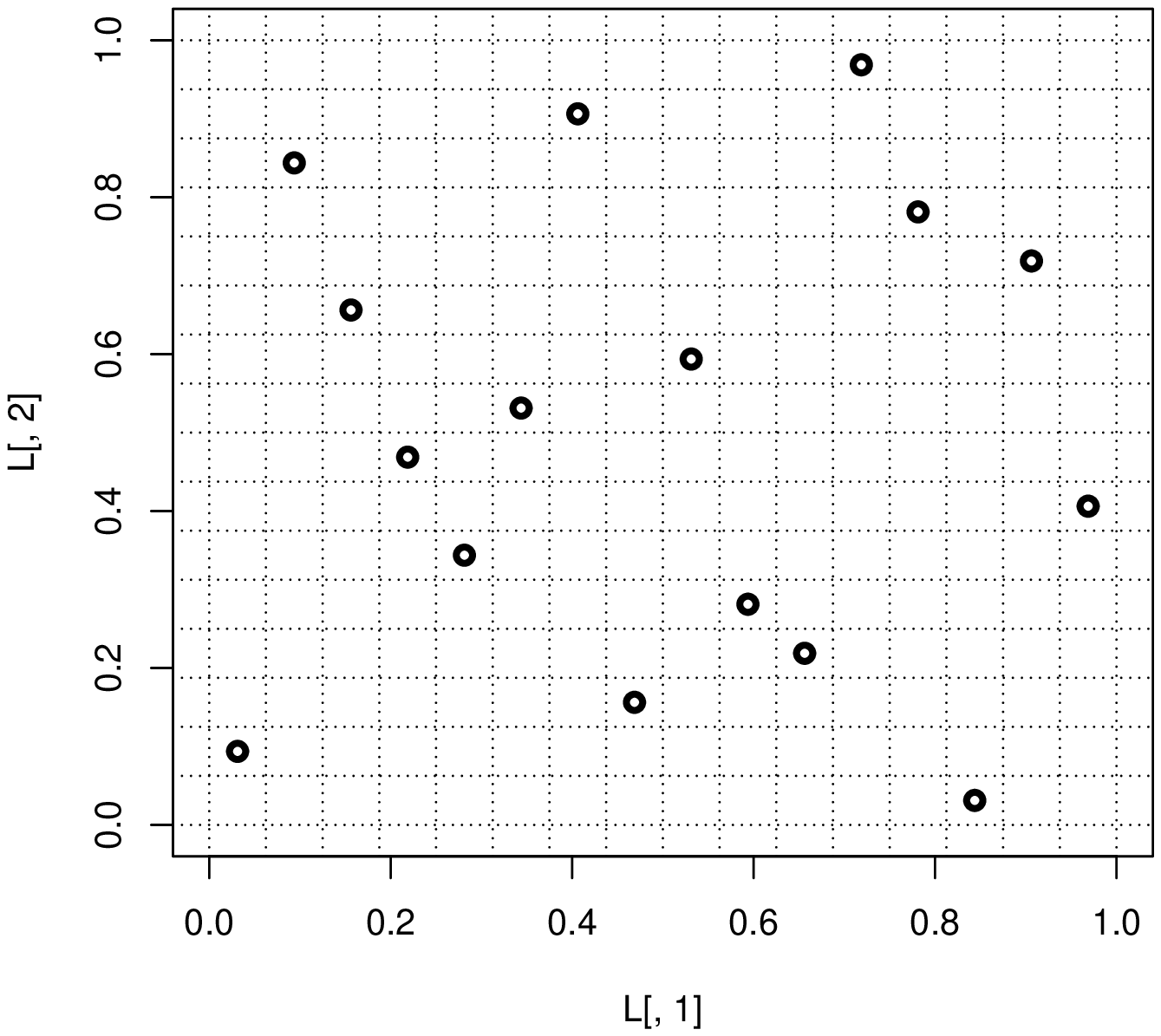}}
\subfigure[A good design ($R_1$ and $R_3$)]{\label{fig:NOA_LHD_G2b} \includegraphics[width=2.0in]{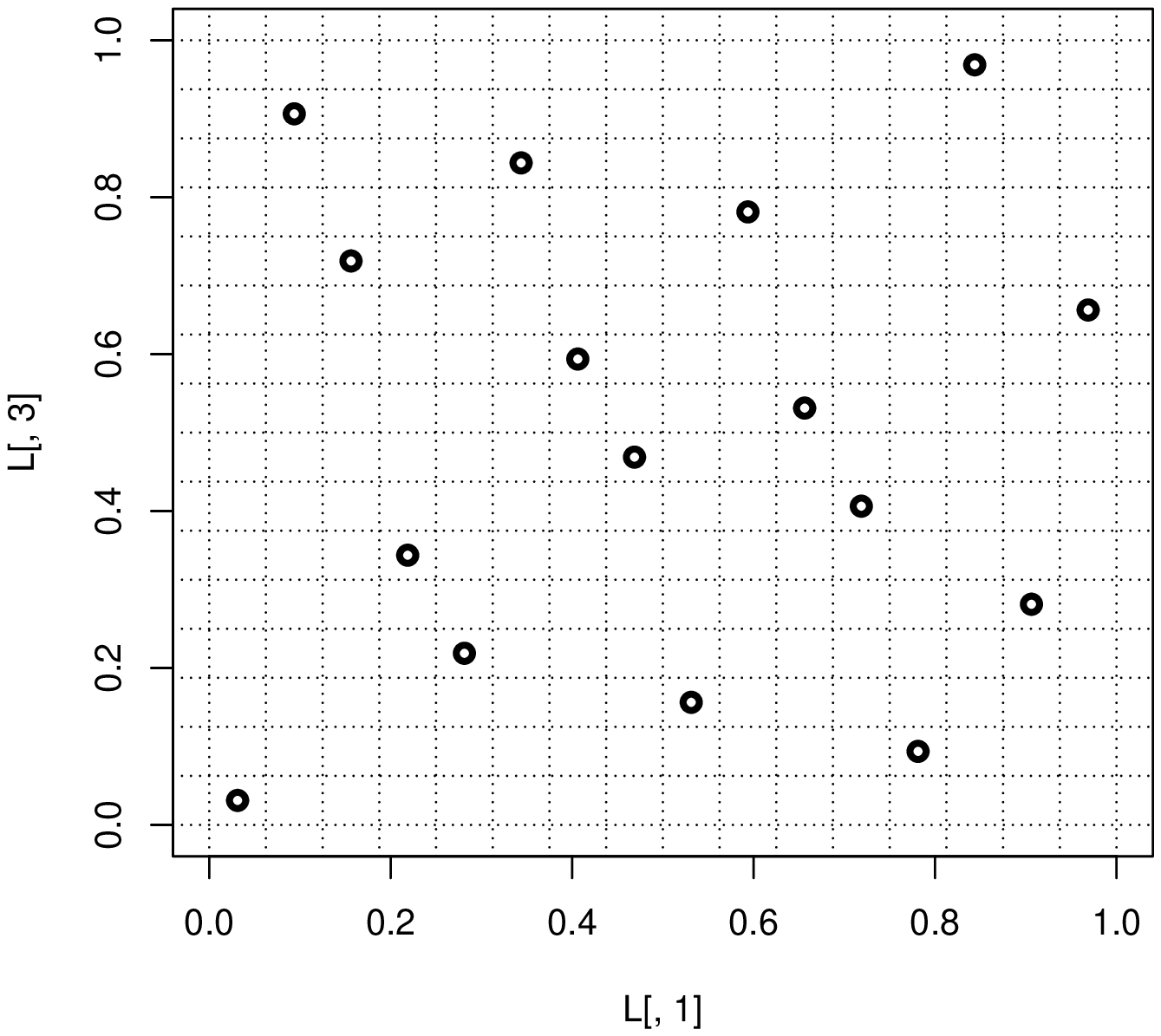}}
\subfigure[A good design ($R_2$ and $R_3$)]{\label{fig:NOA_LHD_G2c} \includegraphics[width=2.0in]{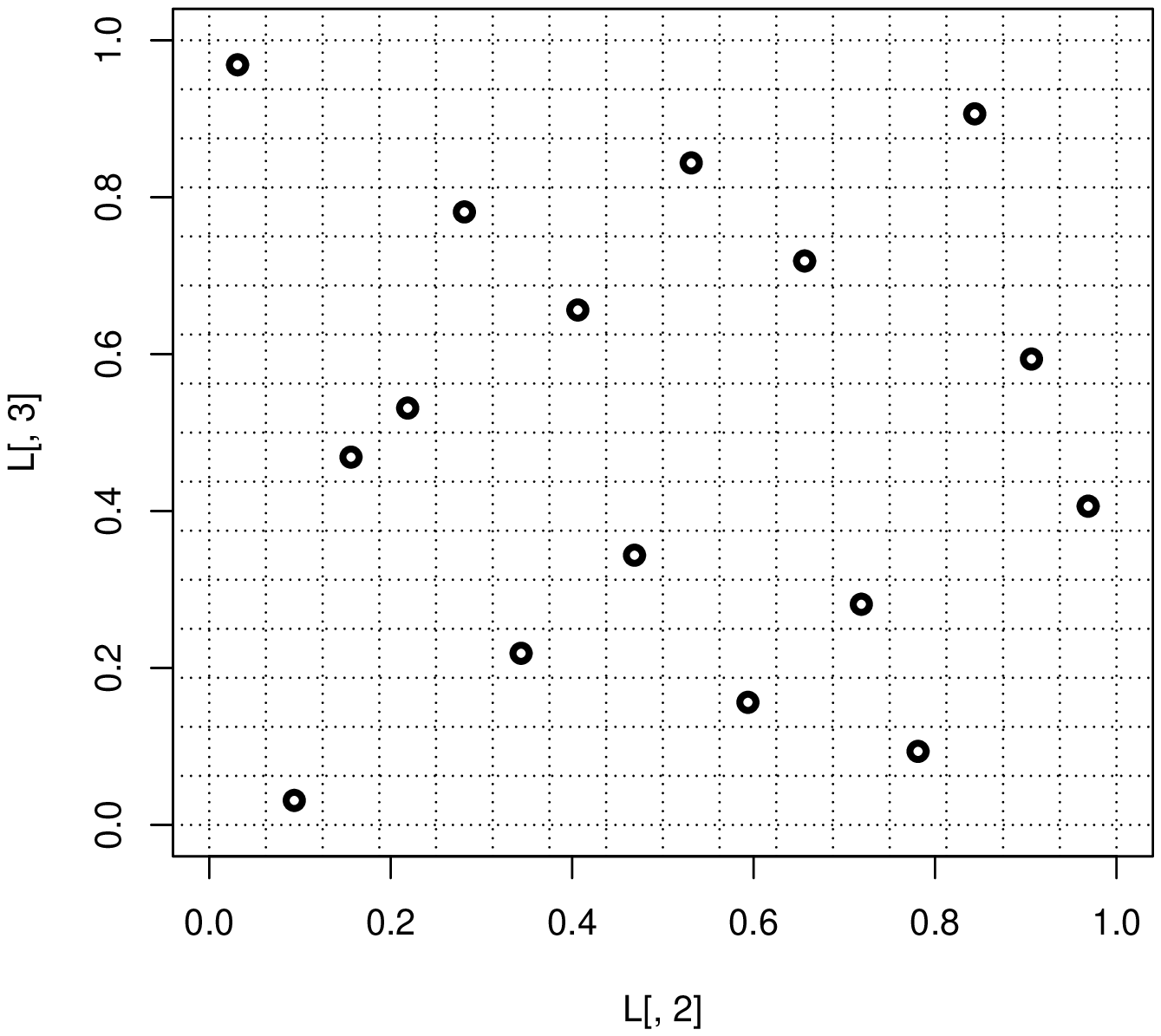}}\vspace{-0.1cm}
\caption{Two-dimensional projections of $NOA(16,8,3,2)$-based LHD in $[0, 1]^3$.}\label{fig:NOA_LHDs_G2}
\end{figure}

A quick glance at Figure~\ref{fig:NOA_LHDs_G2} indicates that all projections are reasonably space-filling, which is supported by the similar AID values (0.5480, 0.5465 and 0.5474 for Figures~\ref{fig:NOA_LHD_G2a}, \ref{fig:NOA_LHD_G2b} and \ref{fig:NOA_LHD_G2c}, respectively). MID values of the two-dimensional projections suggest that the design points in Figure~\ref{fig:NOA_LHD_G2b} is the most spread out (with MID = 0.13975) as compared to the LHDs in Figures~\ref{fig:NOA_LHD_G2a} and \ref{fig:NOA_LHD_G2c} (with MID = 0.08839 for both).
}\end{example}

\begin{example}\label{example:4}{\rm
Consider the same star as in Examples~\ref{example:1} - \ref{example:3}, however, $\delta_l^{(j)}$'s violate only ``G3". Let $R_1 =\langle A,B,ACD \rangle$, $R_2 =\langle C,ABD,ABC \rangle$ and $R_3 =\langle AC,AD,BC \rangle$. Note that $(R_1, R_3)$ violate ``G3" as $\delta_2^{(1)} + \delta_3^{(1)} = B + ACD = ABCD$ and $\delta_2^{(3)} + \delta_3^{(3)} = AD + BC = ABCD$, and $(R_2, R_3)$ violate ``G3" as $\delta_1^{(2)} + \delta_3^{(2)} = C + ABC = AB$ and $\delta_1^{(3)} + \delta_3^{(3)} = AC + BC = AB$. Figure~\ref{fig:NOA_LHDs_G3} shows the two-dimensional LHD projections.

\begin{figure}[h!]\centering
\subfigure[A good design ($R_1$ and $R_2$)]{\label{fig:NOA_LHD_G3a} \includegraphics[width=2.0in]{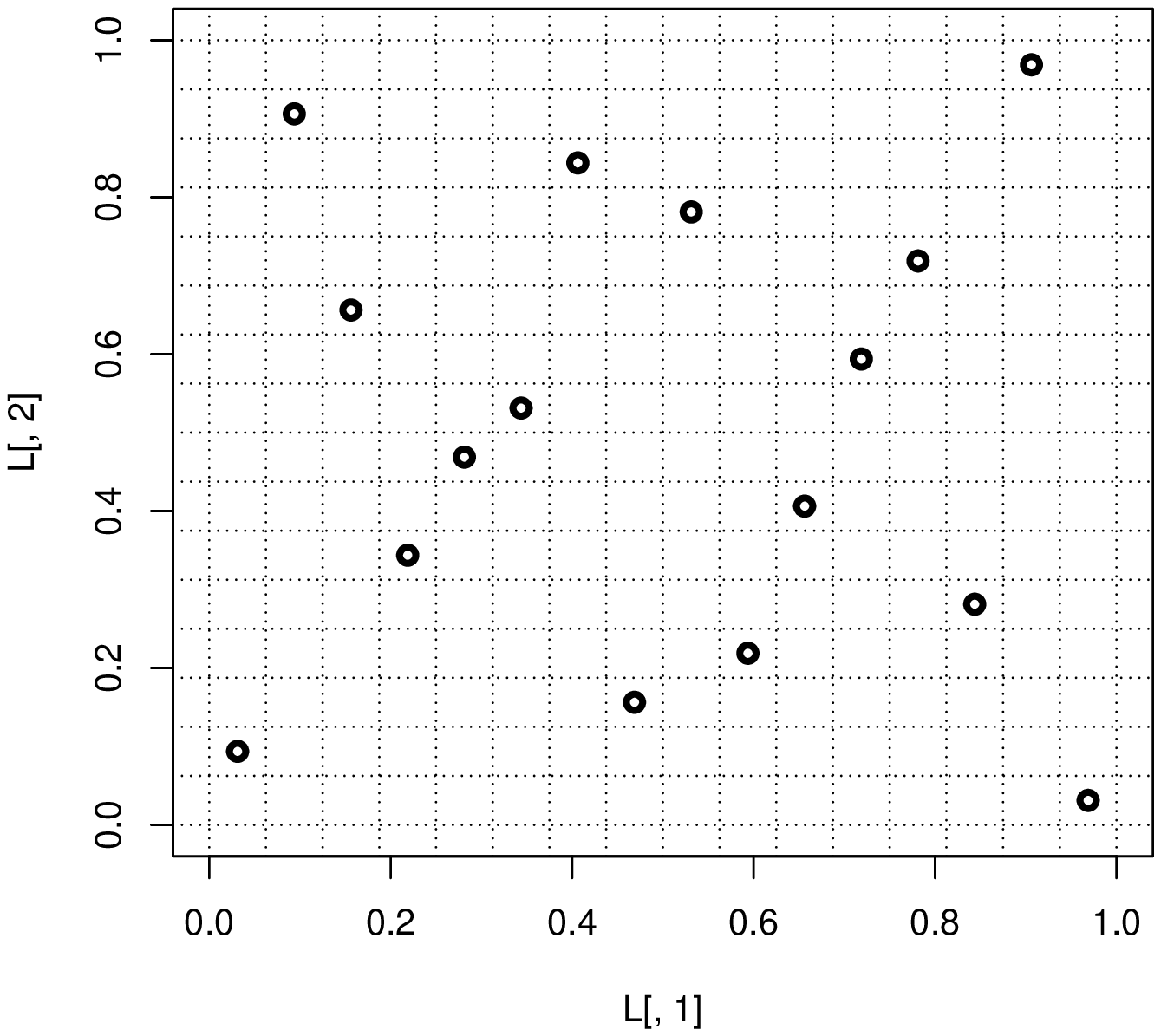}}
\subfigure[A good design ($R_1$ and $R_3$)]{\label{fig:NOA_LHD_G3b} \includegraphics[width=2.0in]{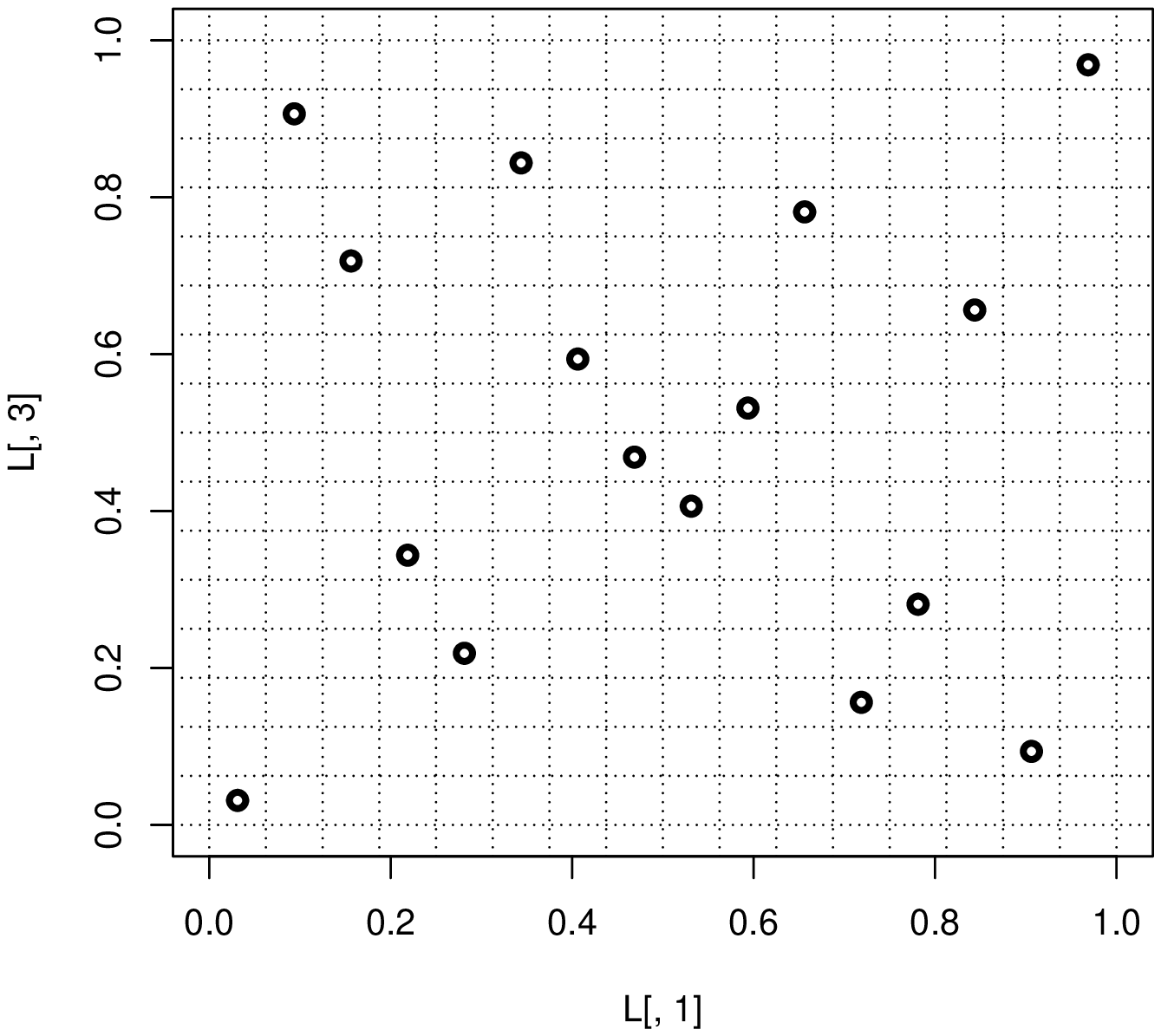}}
\subfigure[A bad design ($R_2$ and $R_3$)]{\label{fig:NOA_LHD_G3c} \includegraphics[width=2.0in]{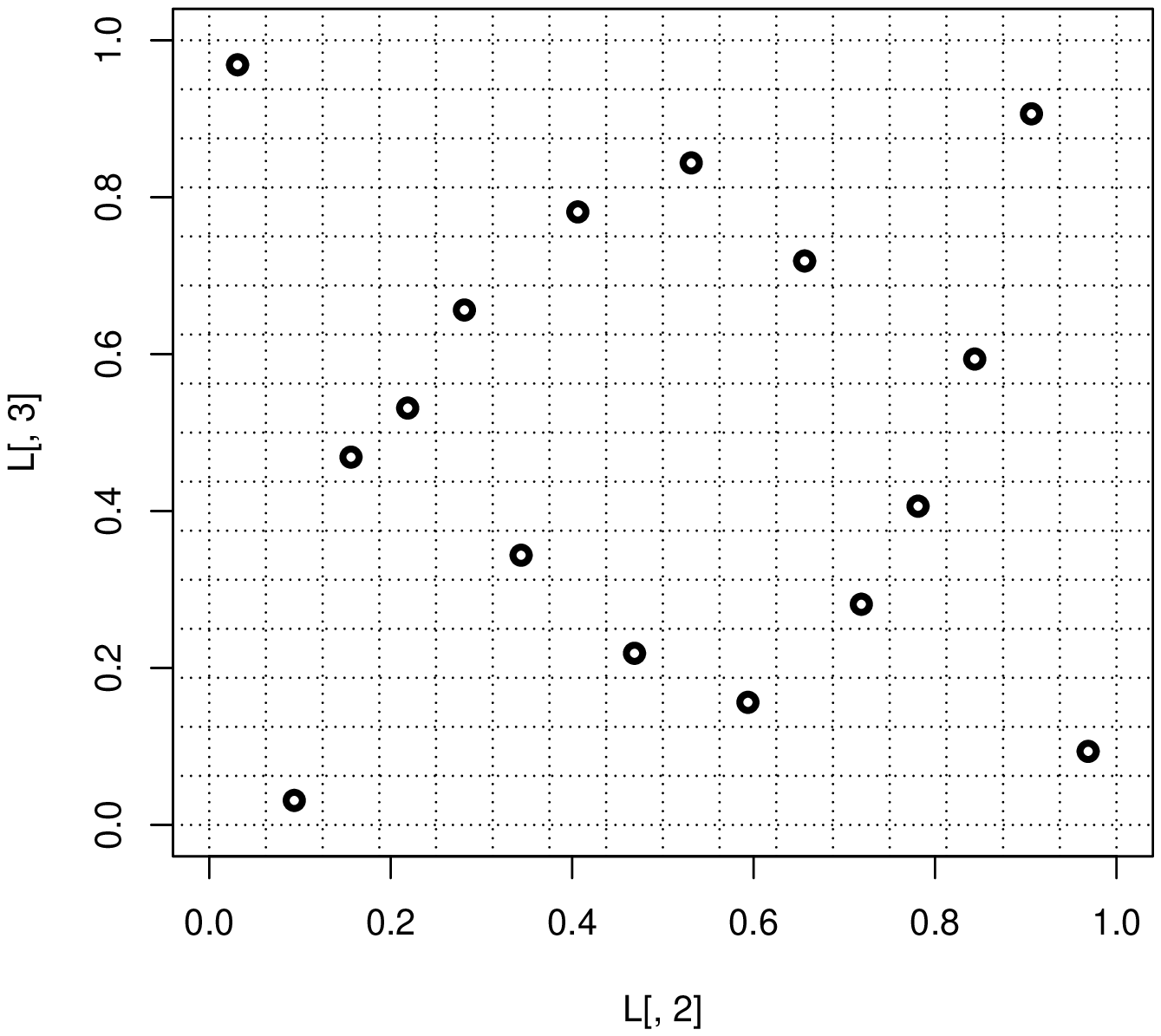}}\vspace{-0.1cm}
\caption{Two-dimensional projections of $NOA(16,8,3,2)$-based LHD in $[0, 1]^3$.}\label{fig:NOA_LHDs_G3}
\end{figure}

It is clear from Figure~\ref{fig:NOA_LHD_G3c} that the violation of ``G3" caused adverse effect on the space-filling behaviour of the LHD. Focusing on the space-filling property of the full three-dimensional LHD, the MID and AID values are 0.1531 and 0.6793 respectively, which are relatively smaller (hence, a worse design) as compared to the designs that violated ``G1" or ``G2" in Examples~\ref{example:2} and \ref{example:3}. All of the two-dimensional projections in Figure~\ref{fig:NOA_LHDs_G3} yield identical MID value of 0.08839, and interestingly, somewhat comparable AID values 0.5441 and 0.5445 for Figures~\ref{fig:NOA_LHD_G3a} and \ref{fig:NOA_LHD_G3c} as well.

}\end{example}

\begin{example}\label{example:5}{\rm
Consider the same star as in Examples~\ref{example:1} - \ref{example:4}, however, we wish to choose $\delta_l^{(j)}$'s that follow all three guidelines (i.e., no violations). Let $R_1 =\langle B,ACD,AB \rangle$, $R_2 =\langle D,C,ABC \rangle$ and $R_3 =\langle AC,BC,CD \rangle$. Note that the set $\{R_1, R_2, R_3\}$ here is very similar to that in Example~\ref{example:2}, except the order of $\delta_l^{(1)}$ (generators of $R_1$) has been changed. Figure~\ref{fig:NOA_LHDs_G0} depicts the corresponding two-dimensional LHD projections.

\begin{figure}[h!]\centering
\subfigure[A good design ($R_1$ and $R_2$)]{\label{fig:NOA_LHD_G0a} \includegraphics[width=2.0in]{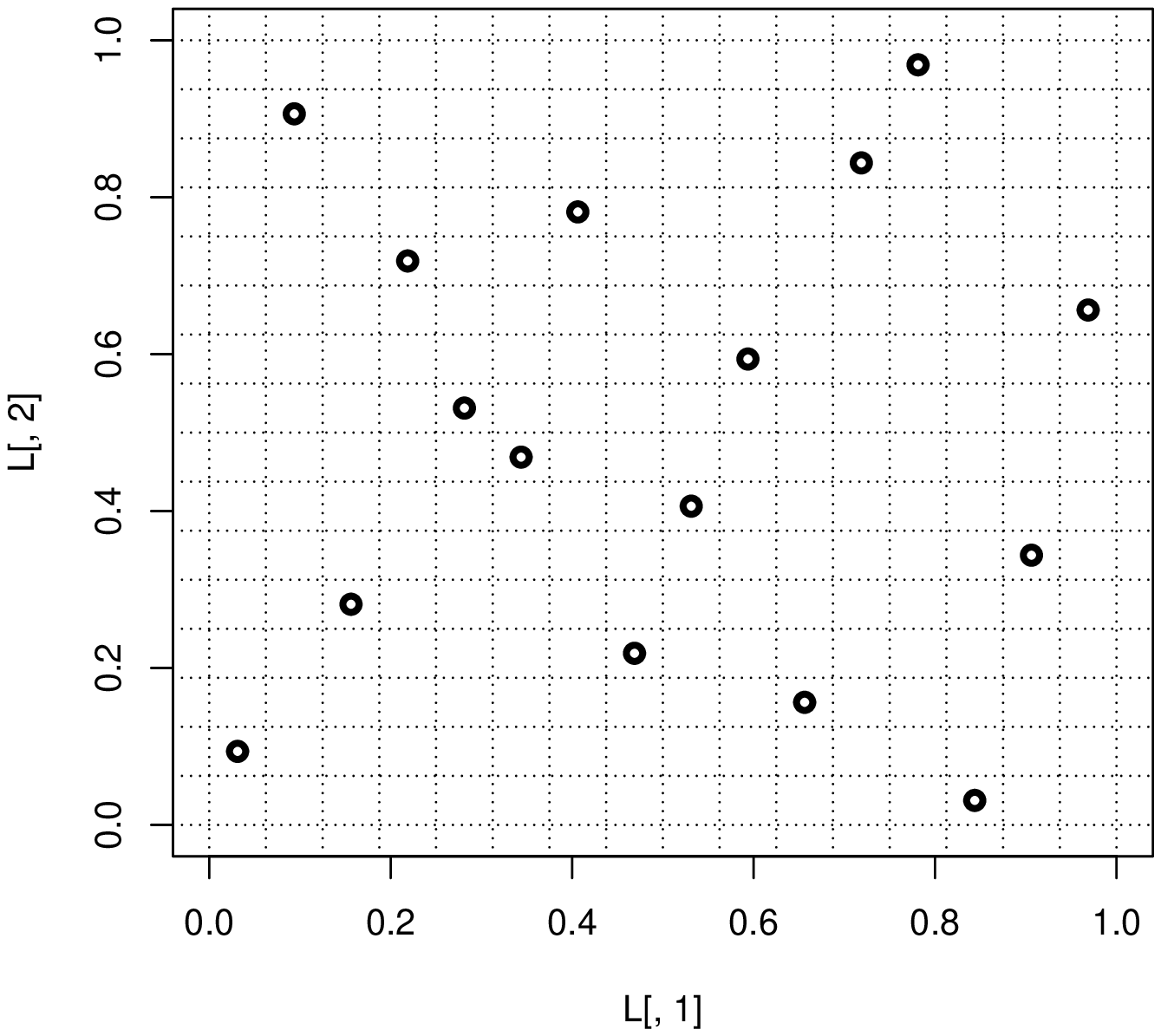}}
\subfigure[A good design ($R_1$ and $R_3$)]{\label{fig:NOA_LHD_G0b} \includegraphics[width=2.0in]{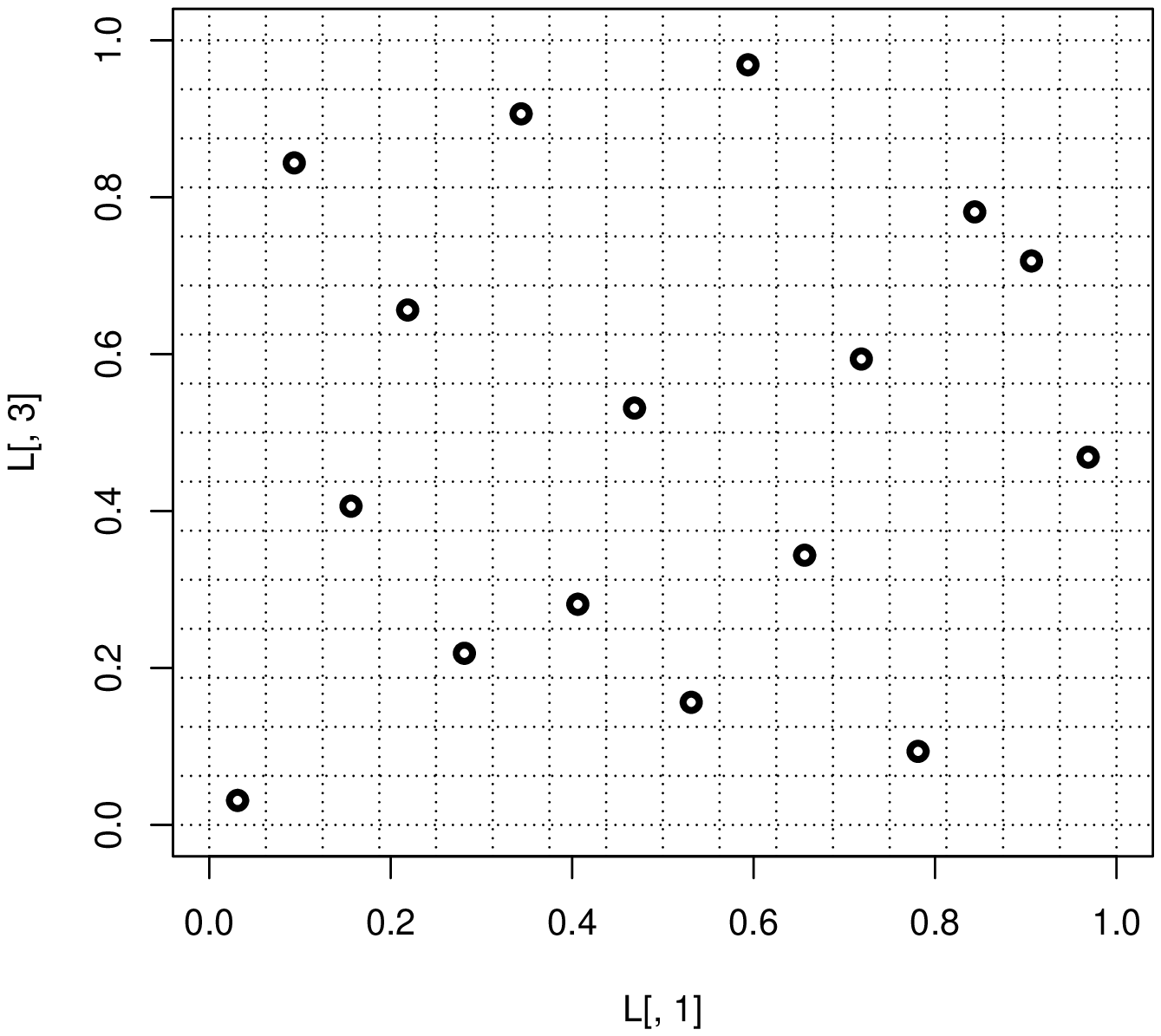}}
\subfigure[A good design ($R_2$ and $R_3$)]{\label{fig:NOA_LHD_G0c} \includegraphics[width=2.0in]{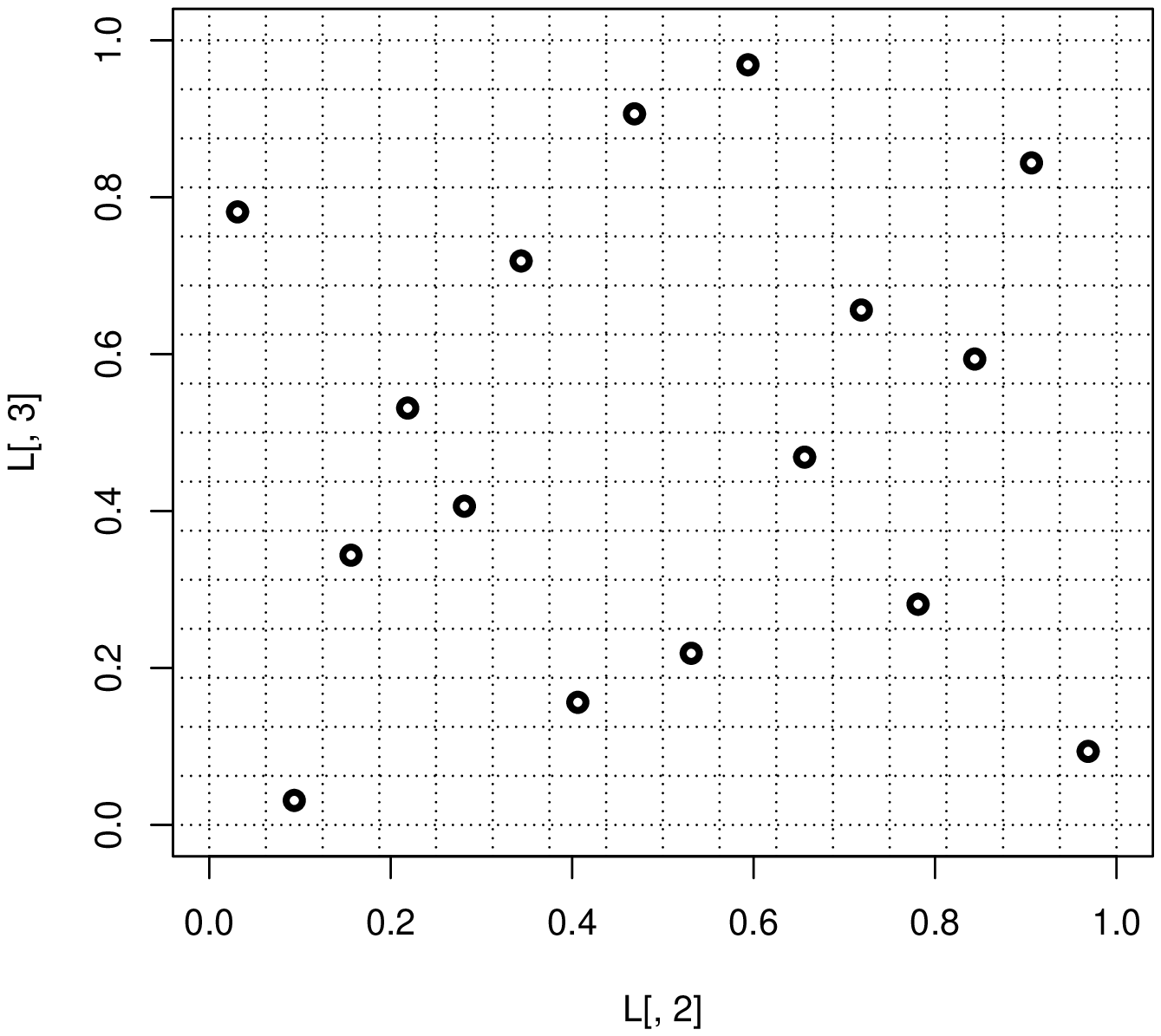}}\vspace{-0.1cm}
\caption{Two-dimensional projections of $NOA(16,8,3,2)$-based LHD in $[0, 1]^3$.}\label{fig:NOA_LHDs_G0}
\end{figure}

A quick glance of Figure~\ref{fig:NOA_LHDs_G0} indicates that the NOA derived from this star generates LHD with space-filling two-dimensional projections. For the full three-dimensional LHD, MID and AID values are 0.2724 and 0.6910 respectively. These values are relatively large compared to the MID and AID values for designs in Examples~\ref{example:2}-\ref{example:4}. Thus, we have obtained a more space-filling LHD by not violating any of the guidelines suggested earlier.


}\end{example}

\begin{figure}[h!]\centering
\includegraphics[width=6in]{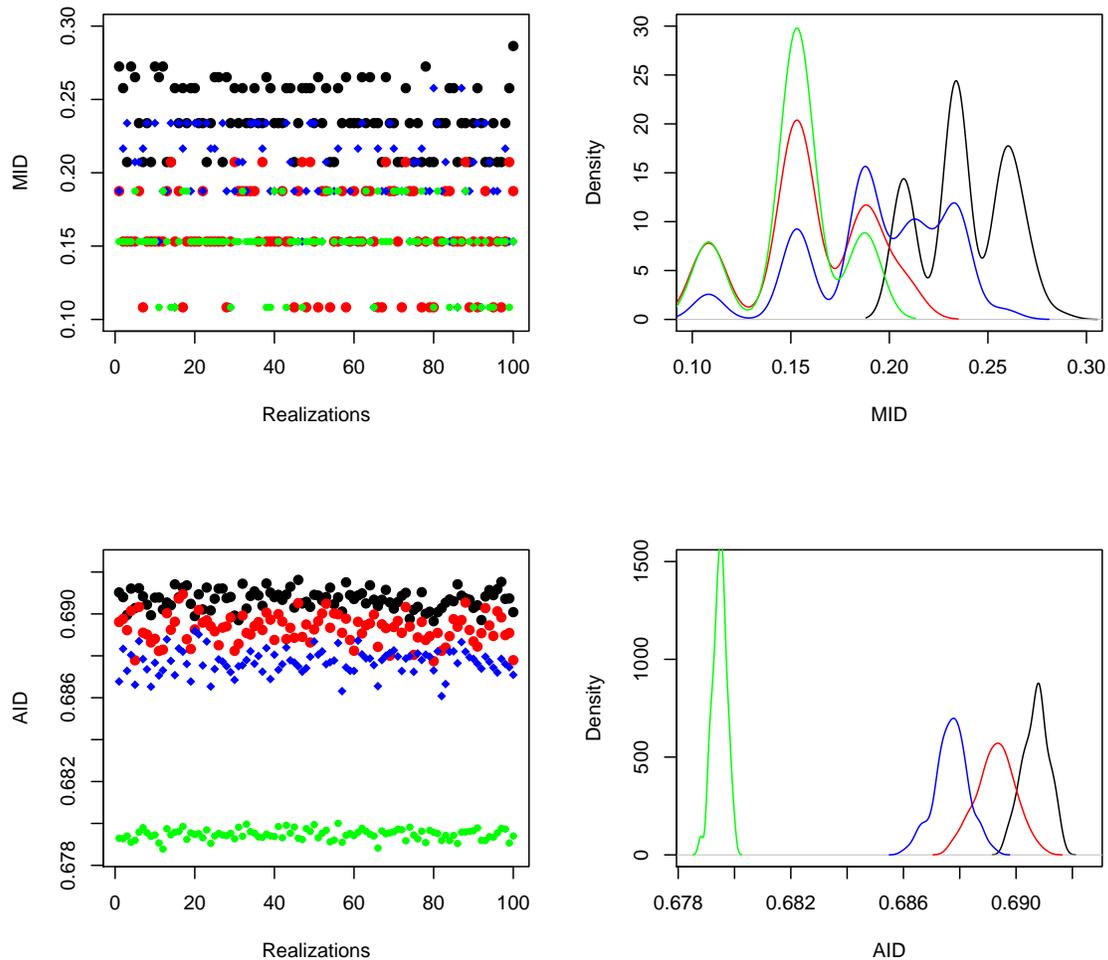}\vspace{-0.5cm}
\caption{[Red - G1 violation; blue - G2 violation; green - G3 violation; black - no violations] Comparison of minimum interpoint distance for $NOA(16,8,3,2)$ shown in Example~1.}\label{fig:simulation-distance}
\end{figure}

For a more comprehensive understanding of the guidelines we conduct a quick simulation study. For each of the four cases (violation of the three guidelines and then no violations) in Examples~2--5, we generate 100 random arrays, $\mathcal{L}$'s, where the randomness is introduced via the permutation of $\{(k-1)n/s_j + 1, ..., kn/s_j\}$. Then for each $\mathcal{L}$, we compute the MID and AID values. Figure~\ref{fig:simulation-distance} compares the dot-plots and densities of these MID and AID values.

It is clear from Figure~\ref{fig:simulation-distance} that the star induced NOA-based LHDs that satisfy the three guidelines are more space-filling (as per MID and AID criteria) that the LHDs that violate any of the three guidelines. Both criteria suggest that violation of G3 is the most damaging, whereas ranking between the damage due to G1 and G2 violations is not unanimous.

\section{Concluding Remarks}\label{sec:conclusion}

This paper presents a new class of space-filling LHDs that are based on geometric NOAs derived from covering stars -- designs for multistage factorial experiments with randomization restrictions. Though we assumed two-level factorial designs (i.e., $PG(p-1,2)$) in Section~3, theoretical results and construction algorithms can easily be generalized for mixed/unbalanced covering or partial stars in $PG(p-1,q)$.

Although we have proposed a few guidelines for constructing space-filling star induced NOA-based LHDs, these guidelines are certainly not exhaustive and somewhat ad-hoc (i.e., not targeted to optimize any criterion like MID or AID). Furthermore, it may not always be possible to satisfy all guidelines (G1, G2 and G3) in a star construction. In such a case, one can use a near orthogonality (Xu and Wu 2001) or space-filling criterion to choose a suitable NOA for constructing space-filling LHD.

In Example~\ref{example:1}, the LHDs are based a covering star $St(3; 3; 2)$ of $PG(3,2)$, i.e., we chose $t=p-1$. One could instead use another $2\le t <p$ for constructing a non-trivial star. For instance, $t=2$ and $t_0=t-1$ would generate a covering star $St(7 ; 2; 1)$ of $PG(3,2)$. The advantage of using $St(7 ; 2; 1)$ instead of $St(3; 3; 2)$ would be the opportunity to construct LHDs with $d=7$, however, the number of levels per factor will be reduced from eight to four. This may have an impact on the space-filling property and is a subject of future research.

\section*{Acknowledgments}

We would like to thank the reviewers for providing helpful comments. We also thank Boxin Tang (Simon Fraser University) for stimulating discussion. Ranjan's work was supported by Discovery grants from the Natural Sciences and Engineering Research Council of Canada.

\begin{center}
{\textbf{REFERENCES}}
\end{center}

\begin{description}

\item {Andr\'e, J.} (1954). Uber nicht-Desarguessche Ebenen mit transitiver Translationsgruppe. \emph{Math. Z.}, 60, 156--186.

\item { Bingham, D., Sitter, R., Kelly, E., Moore, L., \normalfont{and}  Olivas, J. D.} (2008). Factorial designs with multiple levels of randomization. \emph{Statist. Sinica}, {18}, 493--513.

\item {Fang, K.-T., Li, R. \normalfont{and} Sudjianto, A.} (2006). \emph{Design and Modeling for Computer Experiments}. Chapman and Hall/CRC. Boca Raton, FL.

\item {Hedayat, A. S., Sloane, N. J. A. \normalfont{and}  Stufken, J.} (1999). \emph{Orthogonal Arrays: Theory and Applications}. New York: Springer-Verlag.

\item {Iman, R.L. and Conover, W.L.} (1982).  A distribution-free approach to inducing rank correlation among input variables. \emph{Communication in Statistics - Simulation and Computation}, 11, 311--334.

\item { McKay, M. D., Beckman, R. J. \normalfont{and}  Conover, W. J.} (1979), A comparison of three methods for selecting values of input variables in the analysis of output from a computer code, \emph{Technometrics}, {21}(2), 239--245.

\item {Morris, M.D. and Mitchell, T.J.} (1995), Exploratory Designs for Computational Experiments, \emph{Journal of statistical planning and inference}, 43, 381-402.

\item {Nguyen, N-K.} (1996), A note on the construction of near-orthogonal arrays with mixed levels and economic run size. \emph{Technometrics} 38, 279-283.

\item {Owen, A.B.} (1992). Orthogonal arrays for computer experiments, integration and visualization, \emph{Statistical Sinica}, 2, 439-452.

\item { Rains, E.M., Sloane, N.J.A., \normalfont{and}  Stufken, J.} (2002). The lattice of N-run orthogonal arrays. \emph{J. Statist. Plann. Inf.} {102}, 477--500.

\item { Ranjan, P., Bingham, D. \normalfont{and}  Dean, A.} (2009), Existence and Construction of Randomization Defining Contrast Subspaces for Regular Factorial Designs, \textit{The Annals of Statistics}, {37}, 3580 -- 3599.

\item { Ranjan, P., Bingham, D. \normalfont{and}  Mukerjee, R.} (2010), Stars and Regular Fractional Factorial Designs with Randomization Restrictions, \textit{Statist. Sinica}, {20}, 1637-53.

\item { Rasmussen, C. E. \normalfont{and} Williams, C. K. I.} (2006), \emph{Gaussian Processes for Machine Learning}. The MIT Press, Cambridge, MA.

\item { Santner, T.J., Williams, B. \normalfont{and}  Notz, W.} (2003). \emph{The Design and Analysis of Computer Experiments}. Springer Verlag, New York.

\item {Taguchi, G.} (1959), Linear graphs for orthogonal arrays and their applications to experimental designs, with the aid of various techniques. \emph{Report of Statistical Applications Research, Japanese Union of Scientists and Engineers} 6, 1-43.

\item { Tang, B.} (1993). Orthogonal Array-Based Latin Hypercubes. \emph{Journal of the American Statistical Association}, {88}, 1392--1397.

\item {Wang, J. C. and Wu, C. F. J.} (1992), Nearly orthogonal arrays with mixed levels and small runs. \emph{Technometrics}, 34, 409-422.

\item { Wu, C. F. J.  and Hamada, M.} (2000), \emph{Experiments: Planning, Analysis and Parameter Design Optimization}. New York: Wiley.

\item {Xu, H.} (2002), An algorithm for constructing orthogonal and nearly-orthogonal arrays with mixed levels and small runs. \emph{Technometrics}, 44, 356-368.

\item {Xu, H. and Wu, C. F. J.} (2001), Generalized minimum aberration for asymmetrical fractional factorial designs, \emph{The Annals of Statistics}, 29, 1066--1077.

\end{description}

\end{document}